\documentclass{article}
\usepackage{amssymb}

\input{tcilatex}
\begin{document}

\title{Time and energy operators in the canonical quantization of special
relativity}
\author{C.A. Aguill\'{o}n*, M. Bauer** and G.E. Garc\'{\i}a* \\
*Instituto de Ciencias Nucleares, **Instituto de F\'{\i}sica\\
Universidad Nacional Aut\'{o}noma de M\'{e}xico\\
e-mail: bauer@fisica.unam.mx}
\maketitle

\begin{abstract}
Based on Lorentz invariance and Born reciprocity invariance, the canonical
quantization of Special Relativity (SR) is shown to provide a unified origin
for: i) the complex vector space formulation of Quantum Mechanics (QM); ii)
the momentum and space commutation relations and the corresponding
representations; iii) the Dirac Hamiltonian in the formulation of
Relativistic Quantum Mechanics (RQM); iv) the existence of a self adjoint
Time Operator that circumvents Pauli's objection.
\end{abstract}

\section{Introduction}

Quantum mechanics (QM) fails to treat time and space coordinates on the
almost equal footing accorded by Special Relativity (SR) as it does with
momentum and energy. In QM time appears as a parameter not as a dynamical
variable. It is a c-number, following Dirac's designation\cite{Dirac}. This
is the Problem of Time (PoT) in QM, that results in the extensive discussion
of the existence and meaning of a time operator\cite{Muga,Muga2}, and of a
time energy uncertainty relation\cite{Busch,Bauer} in view of Pauli's
objection\cite{Pauli}.

The procedure usually termed "canonical quantization". arises from applying
to the Hamiltonian formulation of classical physics the rule of substituting
dynamical variables by self adjoint operators acting on normalized vectors
representing the physical system. In addition to considering the existance
of Lorentz invariants, the Born reciprocity principle is brought into play%
\cite{Born}. This proposed principle arises from noting that the Hamiltonian
formulation of classical mechanics is invariant under the transformations $%
x_{i}\rightarrow p_{i}$ , $p_{i}\rightarrow -x_{i}$ and from the equivalence
of the configuration and momentum representations in QM. Although Born
acknowledges to be unsuccessful in his intended applications\footnote{%
Born considered the phase space reciprocity invariant $x_{\mu }x^{\mu
}+p_{\mu }p^{\mu }$ as the base to deduce the elemetary particle masses. It
was too early.}, the reciprocity principle is currently receiving\ a renewed
interest\cite{Morgan,Govaerts,Freidel}.In the present paper it is shown that
it complements the required Lorentz invariance in the canonical quantization
of Special Relativity (SR) to provide a unified origin for: i) the complex
vector space formulation of QM; ii) the momentum and position operators'
commutation relations and their corresponding representations; iii) the
Hamiltonian in Dirac's formulation of Relativistic Quantum Mechanics (RQM)%
\cite{Dirac,Thaller,Greiner,Messiah}; iv) the existence of a self adjoint
Time Operator that circumvents Pauli's objection\cite{Bauer1,Bauer2,Bauer3}.

\section{Lorentz and reciprocity invariants in the canonical quantization of
special relativty}

In SR the invariants under Lorentz transformations for a free particle are
the scalar products of the fourvectors in a Minkowski space with metric $%
\eta _{_{^{\mu \nu }}}=diag(1,-1,-1,-1)$\ , namely:%
\begin{equation}
p_{\mu }p^{\mu }=\eta ^{\mu \nu }p_{\mu }p_{\nu }=p_{0}^{2}-\mathbf{p}%
^{2}=(m_{0}c)^{2}\text{ \ \ \ }x_{\mu }x^{\mu }=\eta ^{\mu \nu }x_{\mu
}x_{\nu }=x_{0}^{2}-\mathbf{r}^{2}=s_{0}^{2}
\end{equation}%
where $c$\ is the constant light velocity and\ the constants \ $m_{0}$\ (the
rest mass)\ and \ $s_{0}$\ (to be interpreted)\ are internal properties of
the physical system (Einstein's summation convention is assumed). To be
included in addition are the constant products:%
\begin{equation}
O^{\pm }=x_{\mu }p^{\mu }\pm p^{\mu }x_{\mu }
\end{equation}%
where the symetrization is introduced as these dynamical variables will be
transformed to operators where order matters.

\bigskip

i) \textit{The Dirac free particle Hamiltonian}

From the momentum invariant, first relation in Eq.1, one obtains upon
quantization the QM constraint%
\begin{equation}
\lbrack \hat{p}_{\mu }\hat{p}^{\mu }-(m_{0}c)^{2}]\left\vert \Psi
\right\rangle =0.
\end{equation}%
This can be factorized as%
\begin{equation}
\lbrack \rho ^{\mu }\hat{p}_{\mu }+m_{0}c][\rho ^{\nu }\hat{p}_{\nu
}-m_{0}c]\left\vert \Psi \right\rangle =0,
\end{equation}%
provided that, to cancel the cross terms, the momentum operators satisfy the
commutation relation\ \ $[\hat{p}_{\mu },\hat{p}_{\nu }]=0$ and \ the
coefficients $\rho ^{\mu }\ \ $the\ anticommutation relation$\ \{\rho ^{\mu
}\rho ^{\nu }+\rho ^{\nu }\rho ^{\mu }\}=2\eta ^{\mu \nu }\mathbf{I}_{4}$ \
where $\mathbf{I}_{4}$\ is the $4\times 4$ identity matrix. Thus\ the
coefficients $\ \rho ^{\mu }$\ \ obey a Clifford algebra and are represented
by matrices.

Then the constraint is satisfied with the linear equation:%
\begin{equation}
\lbrack \rho ^{\nu }\hat{p}_{\nu }-m_{0}c]\left\vert \Psi \right\rangle =0.
\end{equation}%
Multiplying by $c\rho ^{0}$\ and defining $\ \rho ^{0}:=\beta ,$ $\ \rho
^{0}\rho ^{i}:=\alpha ^{i}$ one obtains :%
\begin{equation}
c\hat{p}_{0}\left\vert \Psi \right\rangle =\{c\mathbf{\alpha .\hat{p}}+\beta
m_{0}c^{2}\}\left\vert \Psi \right\rangle
\end{equation}%
that exhibits the Dirac Hamiltonian $H_{D}=c\mathbf{\alpha .\hat{p}}+\beta
m_{0}c^{2}$. One recognizes here the procedure followed by Dirac to obtain a
first order linear equation in energy and momentum that agrees with the
second order one resulting from the energy momentum relation in Eq.1\cite%
{Dirac,Thaller,Greiner,Messiah}.

\bigskip

ii) \textit{The free particle time operator}

In exactly the same way, from the second relation in Eq.1, the displacement
invariant yields upon quantization the QM constraint%
\begin{equation}
\lbrack \hat{x}_{\mu }\hat{x}^{\mu }-s_{0}^{2}]\left\vert \Psi \right\rangle
=0.
\end{equation}%
This can be factorized as%
\begin{equation}
\lbrack \rho ^{\mu }\hat{x}_{\mu }+s_{0}][\rho ^{\nu }\hat{x}_{\nu
}-s_{0}]\left\vert \Psi \right\rangle =0
\end{equation}%
provided now\ $[\hat{x}_{\mu },\hat{x}_{\nu }]=0$ and again$\ \ \{\rho ^{\mu
}\rho ^{\nu }+\rho ^{\nu }\rho ^{\mu }\}=2\eta ^{\mu \nu }\mathbf{I}_{4}$.
The constraint is then satisfied with the linear equation:%
\begin{equation}
\lbrack \rho ^{\nu }\hat{x}_{\nu }-s_{0}]\left\vert \Psi \right\rangle =0
\end{equation}%
or, denoting $s_{0}=c\tau _{0}$ where $\tau _{0}$\ would be an internal time
property of the system:%
\begin{equation}
(\hat{x}_{0}/c)\left\vert \Psi \right\rangle =\{\mathbf{\alpha .\hat{r}}%
/c+\beta \tau _{0}\}\left\vert \Psi \right\rangle .
\end{equation}%
Here $T=\mathbf{\alpha .\hat{r}}/c+\beta \tau _{0}$ is the time operator
introduced earlier by analogy to the Dirac Hamiltonian\cite{Bauer2}.

\bigskip

iii) \textit{The Born reciprocity invarian}t

The invariants:%
\begin{equation}
\hat{O}^{\pm }=\{\hat{x}^{\mu }\hat{p}_{\mu }\pm \hat{p}_{\mu }\hat{x}^{\mu
}\}=\{\hat{x}_{0}\hat{p}_{0}-\mathbf{\hat{r}.\hat{p}}\}\pm \{\hat{p}_{0}\hat{%
x}_{0}-\mathbf{\hat{p}.\hat{r}}\}=\{\hat{x}_{0}\hat{p}_{0}\pm \hat{p}_{0}%
\hat{x}_{0}\}-\{\mathbf{\hat{r}.\hat{p}\pm \hat{p}.\hat{r}\}}
\end{equation}%
give the constraint:%
\begin{equation}
\lbrack \{\hat{x}_{0}\hat{p}_{0}\pm \hat{p}_{0}\hat{x}_{0}\}-\{\mathbf{\hat{r%
}.\hat{p}\pm \hat{p}.\hat{r}\}]}\left\vert \Psi \right\rangle =0
\end{equation}%
or equivalently:%
\begin{equation}
\{\hat{x}_{0}\hat{p}_{0}\pm \hat{p}_{0}\hat{x}_{0}\}\left\vert \Psi
\right\rangle =\{\mathbf{\hat{r}.\hat{p}\pm \hat{p}.\hat{r}\}]}\left\vert
\Psi \right\rangle .
\end{equation}

The operator $O^{+}$\ \ is self adjoint and represents a real invariant. On
the other hand $\ O^{-}\ $is a pure imaginary invariant as $\
(O^{-})^{\dagger }=-O^{-}$, but is the one to satisfy reciprocity
invariance. An obvious choice for $O^{-}$\ to satisfy Eqs.12 and 13\ is:%
\begin{equation}
\{\hat{x}_{0}\hat{p}_{0}-\hat{p}_{0}\hat{x}_{0}\}=\{\mathbf{\hat{r}.\hat{p}-%
\hat{p}.\hat{r}\}=}i\hbar
\end{equation}%
where $\hbar $ is the reduced Planck constant.\bigskip

Thus reciprocity invariance complements Lorentz invariance to yield the
commutation relations of the operators $\hat{x}_{\mu }$\ \ and \ $\hat{p}%
_{\nu }$\ , namely:%
\begin{equation}
\lbrack \hat{x}_{\mu },\hat{x}_{\nu }]=0,\ \ \ [\hat{p}_{\mu },\hat{p}_{\nu
}]=0\ \ \ \ \ [\hat{x}_{\mu },\hat{p}_{\nu }]=i\hbar \delta _{\mu \nu }
\end{equation}%
\ as an alternative to the postulate of transforming Poisson brackets to
quantum commutators. In Appendix A it is shown that these commutation
relations are sufficient to derive: a) the continuity from $-\infty $\ \ \
to \ $+\infty $\ of the spectra of $\ \hat{x}_{\mu }$\ \ and \ \ $\hat{p}%
_{\mu }$\ \ ; b) the representations of \ $\hat{x}_{\mu }$\ \ and \ $\hat{p}%
_{\mu }$\ \ in the corresponding orthogonal eigenvector basis: c) the
Fourier transformation between the configuration and momentum representation
of the system vector\ and d) the Heisenberg uncertainty relations, including
Bohr's interpretation of the time-energy uncertainty relation (Appendix B).
Such unified relationship is unfortunately not present in QM textbooks,
where some of these elements are usually introduced as independent\textit{\
antzats}.

\section{Configuration and momentum representations}

Considering Eq.6.in the configuration representation where $\hat{p}_{\nu
}\rightarrow -i\hbar \frac{\partial }{\partial x_{\nu }}$ \ this equation
reads:%
\begin{equation}
i\hbar c\frac{\partial }{\partial x_{0}}\Psi (\mathbf{r,}x_{0})=\{-i\hbar
c\alpha ^{i}\frac{\partial }{\partial x_{i}}+\beta m_{0}c^{2}\}\Psi (\mathbf{%
r,}x_{0}).
\end{equation}%
Substituting from SR $\ x_{0}=ct$ , the result is Dirac's relativistic
equation as usually formulated, namely:%
\begin{equation}
i\hbar \frac{\partial }{\partial t}\Psi (\mathbf{r,}t)=\{-i\hbar c\alpha ^{i}%
\frac{\partial }{\partial x_{i}}+\beta m_{0}c^{2}\}\Psi (\mathbf{r,}t).
\end{equation}

On the other hand, in the momentum representation where $\hat{x}_{\mu
}\rightarrow i\hbar \frac{\partial }{\partial p_{\mu }}$ , Eq.10 yields:%
\begin{equation}
i\hbar \frac{\partial }{\partial cp_{0}}\Phi (\mathbf{p},p_{0})=\{(i\hbar
/c)\alpha ^{i}\frac{\partial }{\partial p_{i}}+\beta \tau _{0}\}\Phi (%
\mathbf{p},p_{0}).
\end{equation}%
Substituting \ $cp_{0}=e$, one obtains for the time operator the equation:%
\begin{equation}
i\hbar \frac{\partial }{\partial e}\Phi (\mathbf{p},e)=\{(i\hbar /c)\alpha
^{i}\frac{\partial }{\partial p_{i}}+\beta \tau _{0}\}\Phi (\mathbf{p},e)
\end{equation}

\section{The energy and time spectra, and Pauli's objection}

As is well known, the energy spectrum of the Dirac Hamiltonian has both
positive and negative real values, namely $e(p)=\pm \sqrt{(cp)^{2}+m_{0}c^{2}%
}$ ,.separated by a $2m_{0}c^{2}$\ gap. In the same way, the spectrum of the
time operator contains positive and negative real values separated by a gap $%
2\tau _{0}$, as $\tau (r)=\pm \sqrt{(r/c)^{2}+\tau _{0}}$.

Now, the actual interpretation of the effect of the Dirac Hamiltonian $\
H_{D}$\ and the time operator $\ T$\ \ is seen from the fact that they\ are
self adjoint. By Stone-vonNewmann's theorem\cite{Jordan} they are generators
of unitary transformations of the state vectors. Then it can be shown that
for infinitesimal changes\cite{Bauer3}:

a) \ $U_{T}=\exp \{i(\delta e)T/\hbar \}\thickapprox \exp \{i(\delta e)%
\mathbf{\alpha .r}/c\hbar \}\exp \{i(\delta e)\beta \tau _{0}\}$ generates
continous displacements in momentum $\delta \mathbf{p}=(\mathbf{\alpha /}c%
\mathbf{)}\delta e=$ $c\mathbf{\alpha (\delta e}/c^{2})$ and changes in
phase $\delta \phi =\beta \tau _{0}\delta e/\hbar $. A $2\pi $ finite phase
change for positive energy waves ($\left\langle \beta \right\rangle =1$) is
obtained from setting:%
\begin{equation}
\tau _{0}=h/m_{0}c^{2}=T_{B}\text{ \ \ \ \ \ }\Delta e=m_{0}c^{2}=h/T_{B}%
\text{\ }
\end{equation}%
Then $\tau _{0}$\ is seen to be the deBroglie period $T_{B}$, in agreement
with deBroglie's daring assumptions\cite{Broglie,Baylis}. It is an intrinsic
time property associated with the rest mass\cite{Lan}.

b) \ $U_{H_{D}}=\exp \{i(\delta t)H_{D}/\hbar \}$ $\thickapprox \exp
\{i(\delta t)c\mathbf{\alpha .p}/\hbar \}\exp \{i(\delta t)\beta
m_{0}c^{2}\} $ generates displacements in space $\delta \mathbf{r}=c\mathbf{%
\alpha \delta }t$ and changes in phase $\delta \phi =\beta m_{0}c^{2}\delta
t/\hbar $. For $\left\langle \beta \right\rangle =1$, a $2\pi $ finite
change of phase requires a time lapse:%
\begin{equation}
\Delta t=2\pi \hbar /m_{0}c^{2}=h/m_{0}c^{2}=T_{B}
\end{equation}

For wave packets the expectation value\ $\left\langle c\mathbf{\alpha }%
\right\rangle $ is the group velocity $\mathbf{v}_{gp}$ and the space
displacement in a time lapse $\Delta t$\ generated by $H_{D}$\ \ corresponds
to the classical $\mathbf{v}_{gp}\Delta t$. On the other hand $T$\ \ acts on
the continous momentum space, generating a change of momentum $\Delta 
\mathbf{p}=m\mathbf{v}_{gp}=m_{0}\gamma \mathbf{v}_{gp}$, where $\gamma
=[1-(v/c)^{2}]^{-1}$ is the Lorentz factor, and consequently an energy
change from $E(\mathbf{p})$ to $E(\mathbf{p}+m\mathbf{v}_{gp})$ in both
branches of the relativistic energy spectrum. This circumvents Pauli's
correct objection that a commutation relation $[T,H]=i\hbar $ where $T$ acts
on the energy spectrum, necessarily implies a continuum energy spectrum from 
$-\infty $ to $+\infty $ , contradicting the fact that the energy
expectation value is expected to be positive and that there also may be
discrete eigenvalues\cite{Pauli}.

To be remarked finally is that, from Eq.20,\ the energy gap $2m_{0}c^{2}$\
and the time gap\ $2h/m_{0}c^{2}$\ are complementary of each other. The mass
dependence of the Zitterbewegung period (twice the deBroglie period) has
been correctly exhibited in the experimental simulation of the Dirac
equation with trapped ions\cite{Gerritsma,Bauer4}.

\section{Conclusion}

It has been shown that the canonical quantization of SR that preserves the
Lorentz and reciprocity invariants, is at the origin of the (usually
postulated or inferred separately) commutation relations of the
configuration and momentum dynamical operators, as well as of the Dirac
relativistic Hamiltonian together with a self adjoint relativistic "time
operator". Furthermore, it brings about the derived properties - infinite
continuous space and momentum spectra, ensuing representations, uncertainty
relation as shown in Appendix A and B, that unfortunately in most QM
textbooks are introduced as independent\textit{\ antzats}.

To be stressed also, it is the reciprocity invariance which introduces an
imaginary constant, opening the formulation to complex functions which are
necessary to allow for "a non-negative probability function that is constant
in time when integrated over the whole space "\cite{Pauli}, the basis for a
probabilistic interpretation of QM.

The problem of time is very much present in the canonical quantization of
General Relativity (GR), with many facets: indefinition of the spacetime
foliation ("many fingered time"), disappearence of time ("frozen
formalism"), and so on\cite{Isham,Kuchar,Butterfield,Anderson2}. However one
condition to be satisfied is local concordance with SR, i.e., any acceptable
theory of Quantum Gravity (QG) must allow to recover the classical spacetime
in the appropiate limit\cite{Bonder}. It follows that a venue to be explored
is whether this bottom up completion of Dirac's RQM with a time operator as
derived above helps to resolve some of the issues noted\cite{Bauer4,Bauer5}%
.

\section{{\protect\Large Appendix A. The lore of \ \ }$[\hat{x},\hat{p}%
]=i\hbar $}

To represent observables the operators $\hat{x}_{\mu }$ and $\hat{p}_{\mu }$
are selfadjoint $(\hat{x}_{\mu }=\hat{x}_{\mu }^{\dagger }$ , $\hat{p}_{\mu
}=\hat{p}_{\mu }^{\dagger })$ , which insures real eigenvalues. Then, for
each component $\ \hat{x}_{\mu }$ and $\ \hat{p}_{\mu }$ it follows:

\bigskip

\textbf{1) Spectrum\cite{Messiah}}

Consider the eigenvalue equation :%
\begin{equation}
\hat{x}\ \left\vert x\right\rangle =x\ \left\vert x\right\rangle  \tag{A.1}
\end{equation}%
By Stone-von Newmann's theorem the operator $U(\alpha )=\exp (-i\alpha \hat{p%
}/\hslash )$ with $a$ real is unitary\cite{Jordan}. Then it can be shown
that:%
\begin{equation}
\hat{x}\ \{U(\alpha )\left\vert x\right\rangle \}=(x+\alpha )\{U(\alpha
)\left\vert x\right\rangle \}  \tag{A.2}
\end{equation}%
Therefore $\{U(\alpha )\left\vert x\right\rangle \}=C\left\vert x+\alpha
\right\rangle $. As $\alpha $\ is arbitrary, it follows that the eigenvalues
of $\hat{x}\ $are continous from\ $-\infty $\ to $+\infty $\ , and that the
eigenvectors satisfy:%
\begin{equation}
\left\langle x^{\prime }\mid x\right\rangle =\delta (x^{\prime }-x)\ \ \ \ \
\ \ \ \ \ \ \int dx\ \left\vert x\right\rangle \left\langle x\right\vert =I 
\tag{A.3}
\end{equation}%
where $\delta (x^{\prime }-x)$\ \ \ is the Dirac delta function and $I$ is
the identity operator.

In the same way one can prove that the eigenvalues in \ $\hat{p}\ \left\vert
p\right\rangle =p\ \left\vert p\right\rangle $\ \ span a continuum from $%
-\infty $\ to $+\infty $\ and that the eigenvectors satisfy:%
\begin{equation}
\left\langle p^{\prime }\mid p\right\rangle =\delta (p^{\prime }-p)\ \ \ \ \
\ \ \ \ \ \ \int dp\ \left\vert p\right\rangle \left\langle p\right\vert =I 
\tag{A.4}
\end{equation}

\textbf{2) Representations}

From Eq.A.3:%
\[
\left\langle x^{\prime }\left\vert \hat{x}\right\vert x\right\rangle
=x\delta (x^{\prime }-x) 
\]%
and%
\begin{eqnarray*}
\left\langle x^{\prime }\right\vert \left[ \hat{x},\hat{p}\right] \left\vert
x\right\rangle &=&i\hslash \delta (x^{\prime }-x)=\left\langle x^{\prime
}\right\vert \hat{x}\hat{p}-\hat{p}\hat{x}\left\vert x\right\rangle \\
&=&x^{\prime }\left\langle x^{\prime }\right\vert \hat{p}\left\vert
x\right\rangle -x\left\langle x^{\prime }\right\vert \hat{p}\left\vert
x\right\rangle =(x^{\prime }-x)\left\langle x^{\prime }\right\vert \hat{p}%
\left\vert x\right\rangle
\end{eqnarray*}%
It follows:%
\begin{equation}
\left\langle x^{\prime }\right\vert \hat{p}\left\vert x\right\rangle =\frac{%
i\hslash \delta (x^{\prime }-x)}{(x^{\prime }-x)}\Longrightarrow _{x^{\prime
}\rightarrow x}i\hslash \frac{d}{dx^{\prime }}\delta (x^{\prime }-x) 
\tag{A,5}
\end{equation}

Introducing the vectors:%
\[
\left\vert \Theta \right\rangle =\hat{x}\left\vert \Psi \right\rangle \text{
\ \ \ \ }\left\vert \Phi \right\rangle =\hat{p}\left\vert \Psi \right\rangle 
\text{\ } 
\]%
their representations in configuration space are:%
\begin{equation}
\Theta (x)=\left\langle x\right\vert \hat{x}\left\vert \Psi \right\rangle
=x\left\langle x\mid \Psi \right\rangle =x\Psi (x)  \tag{A.6}
\end{equation}%
and using Eq.A.5:%
\begin{eqnarray}
\Phi (x) &=&\left\langle x\mid \Phi \right\rangle =\left\langle x\mid \hat{p}%
\mid \Psi \right\rangle =  \nonumber \\
&=&\int dx^{\prime }\left\langle x\right\vert \hat{p}\left\vert x^{\prime
}\right\rangle \left\langle x^{\prime }\mid \Psi \right\rangle =i\hslash
\int dx^{\prime }[\frac{d}{dx^{\prime }}\delta (x^{\prime }-x)]\left\langle
x^{\prime }\mid \Psi \right\rangle  \TCItag{A.7} \\
&=&\left[ \delta (x^{\prime }-x)\Psi (x^{\prime })\right] _{-\infty
}^{+\infty }-i\hslash \int dx^{\prime }\delta (x^{\prime }-x)\frac{d}{%
dx^{\prime }}\Psi (x^{\prime })=-i\hslash \frac{d}{dx}\Psi (x)  \nonumber
\end{eqnarray}%
i.e., the representation in configuration space of the vector $\left\vert
\Phi \right\rangle =\hat{p}\left\vert \Psi \right\rangle $\ is obtained by
taking the derivative of the representation of the vector $\left\vert \Psi
\right\rangle $, while the representation in configuration space of the
vector $\left\vert \Theta \right\rangle =\hat{x}\left\vert \Psi
\right\rangle $ is obtained multiplying by \ $x$\ the representation of $%
\left\vert \Psi \right\rangle $.

To conclude, in configuration space one has:%
\begin{equation}
\hat{x}\Longrightarrow x\ \ \ \ \ \ \ \ \ \ \ \ \ \hat{p}\Longrightarrow
-i\hslash \frac{d}{dx}  \tag{A.8}
\end{equation}%
In the same way in momentum space:%
\begin{equation}
\hat{x}\Longrightarrow i\hslash \frac{d}{dp}\ \ \ \ \ \ \ \ \ \ \ \ \ \hat{p}%
\Longrightarrow \ p  \tag{A.9}
\end{equation}

3) \textbf{Transformation between representations}

Consider:%
\[
\left\langle x\mid \left[ \hat{x},\hat{p}\right] \mid p\right\rangle
=i\hslash \left\langle x\mid p\right\rangle 
\]%
Developing:%
\begin{eqnarray*}
\left\langle x\mid \left[ \hat{x},\hat{p}\right] \mid p\right\rangle
&=&\left\langle x\mid \hat{x}\hat{p}\mid p\right\rangle -\left\langle x\mid 
\hat{p}\hat{x}\mid p\right\rangle \\
&=&xp\left\langle x\mid p\right\rangle -\int dx^{\prime }\left\langle x\mid 
\hat{p}\mid x^{\prime }\right\rangle \left\langle x^{\prime }\mid \hat{x}%
\mid p\right\rangle \\
&=&xp\left\langle x\mid p\right\rangle -i\hslash \int dx^{\prime }[\frac{d}{%
dx^{\prime }}\delta (x^{\prime }-x)]x^{\prime }\left\langle x^{\prime }\mid
p\right\rangle \\
&=&xp\left\langle x\mid p\right\rangle +i\hslash \lbrack \left\langle x\mid
p\right\rangle +i\hslash x\frac{d}{dx}\left\langle x\mid p\right\rangle ]
\end{eqnarray*}%
one obtains:%
\[
xp\left\langle x\mid p\right\rangle +i\hslash \lbrack \left\langle x\mid
p\right\rangle +i\hslash x\frac{d}{dx}\left\langle x\mid p\right\rangle
]=i\hslash \left\langle x\mid p\right\rangle 
\]%
Thus:%
\begin{equation}
i\hslash \frac{d}{dx}\left\langle x\mid p\right\rangle =-p\left\langle x\mid
p\right\rangle  \tag{A.10}
\end{equation}%
which is satisfied if:%
\begin{equation}
\left\langle x\mid p\right\rangle =Ce^{ipx/\hslash }\ \ \ \ \ \ \ \ \ \
\left\langle p\mid x\right\rangle =C^{\ast }e^{-ipx/\hslash }  \tag{A.11}
\end{equation}

Finally:%
\begin{equation}
\Phi (p)=\left\langle p\mid \Psi \right\rangle =\int dx\left\langle p\mid
x\right\rangle \left\langle x\mid \Psi \right\rangle =C^{\ast }\int dx\
e^{-ipx/\hslash }\ \Psi (x)  \tag{A.12}
\end{equation}%
and:%
\begin{equation}
\Psi (x)=\left\langle x\mid \Psi \right\rangle =\int dp\left\langle x\mid
p\right\rangle \left\langle p\mid \Psi \right\rangle =C\int dx\
e^{ipx/\hslash }\ \Phi (p)  \tag{A.13}
\end{equation}
i.e., the representations of the state vector in the configuration and
momentum spaces are\textit{\ Fourier transforms }of each\textit{\ }other%
\textit{.} To preserve normalization one requires $C=C^{\ast }=1/\sqrt{2\pi
\hslash }$.

\bigskip

4) \textbf{Uncertainty relation}

Consider the state vectors%
\begin{equation}
\left\vert \Phi \right\rangle =(\hat{x}-\left\langle x\right\rangle
)\left\vert \Psi \right\rangle \ \ \ \ and\ \ \left\vert \Xi \right\rangle =(%
\hat{p}-\left\langle p\right\rangle )\left\vert \Psi \right\rangle 
\tag{A.14}
\end{equation}

Then%
\begin{equation}
\left\langle \Phi \mid \Phi \right\rangle =\left\langle \Psi \right\vert 
\hat{x}^{2}\left\vert \Psi \right\rangle -\left\langle \Psi \right\vert \hat{%
x}\left\vert \Psi \right\rangle ^{2}=(\Delta x)_{\Psi }^{2}.....\left\langle
\Xi \mid \Xi \right\rangle =\left\langle \Psi \right\vert \hat{p}%
^{2}\left\vert \Psi \right\rangle -\left\langle \Psi \right\vert \hat{p}%
\left\vert \Psi \right\rangle ^{2}=(\Delta p)_{\Psi }^{2}  \tag{A.15}
\end{equation}

By Schawrz inequality one has%
\begin{eqnarray*}
\left\langle \Phi \mid \Phi \right\rangle \left\langle \Xi \mid \Xi
\right\rangle &\geq &\left\vert \left\langle \Phi \mid \Xi \right\rangle
\right\vert ^{2}= \\
&=&\left\vert \left\langle \Psi \right\vert 
{\frac12}%
[\hat{x},\hat{p}]+%
{\frac12}%
\{\hat{x},\hat{p}\}-\left\langle x\right\rangle \left\langle p\right\rangle
\left\vert \Psi \right\rangle \right\vert ^{2}\geq \\
&\geq &\left\vert \left\langle \Psi \right\vert 
{\frac12}%
[\hat{x},\hat{p}]\left\vert \Psi \right\rangle \right\vert ^{2}=(\hslash
/2)^{2}
\end{eqnarray*}

Finally%
\begin{equation}
(\Delta x)_{\Psi }(\Delta p)_{\Psi }\geq \hslash /2  \tag{A.16}
\end{equation}

\bigskip

\section{Appendix B. The time-energy uncertainty relation}

The Dirac Hamiltonian and the time operator satisfy the commutation relation%
\begin{equation}
\lbrack T,H_{D}]=i\hbar \{I+2\beta K\}+2\beta \{\tau _{0}H_{D}-m_{0}c^{2}T\}
\tag{B.1}
\end{equation}%
where $K=\beta (2\mathbf{s.l}/\hbar ^{2}+1)$\ is a constant of motion\cite%
{Thaller}. In the usual manner an uncertainty relation follows, namely:%
\begin{equation}
(\Delta T)(\Delta H_{D})\geq (\hbar /2)\left\vert \{1+2<\beta
K>\}\right\vert =(\hbar /2)\left\vert \{3+4\left\langle \mathbf{s.l}/\hbar
^{2}\right\rangle \}\right\vert  \tag{B.2}
\end{equation}

Consider:%
\begin{eqnarray}
(\Delta T)^{2} &=&\left\langle T^{2}\right\rangle -\left\langle
T\right\rangle ^{2}=\left\langle r^{2}/c^{2}+\tau _{0}^{2}\right\rangle
-\left\langle \mathbf{\alpha .r}/c+\beta \tau _{0}\right\rangle ^{2}= 
\TCItag{B.3} \\
&=&(1/c^{2})(\Delta r)^{2}+[\left\langle r/c\right\rangle ^{2}+\tau
_{0}^{2}]-[\left\langle \mathbf{\alpha .r}/c+\beta \tau _{0}\right\rangle
^{2}]\gtrapprox (1/c^{2})(\Delta r)^{2}  \nonumber
\end{eqnarray}%
In the same way:%
\begin{equation}
(1/c^{2})(\Delta H_{D})^{2}\gtrapprox (\Delta p)^{2}  \tag{B.4}
\end{equation}%
It follows finally:%
\begin{equation}
(\Delta T)^{2}(\Delta H_{D})^{2}\gtrapprox (\Delta r)^{2}(\Delta p)^{2} 
\tag{B.5}
\end{equation}%
This correspponds to Bohr's interpretation that the uncertainty in the time
of passage at a certain point is given by the width of the wave packet,
which is complementary to the momentum uncertainty, and thus to the energy
uncertainty\cite{Bohr}.

\end{document}